# *Ultra-fast (milliseconds), multi-dimensional RF pulse design with deep learning*

Mads Sloth Vinding[1], Birk Skyum[2], Ryan Sangill[1], Torben Ellegaard Lund[1]

[1]Center of Functionally Integrative Neuroscience, Aarhus University, Denmark
[2]Interdisciplinary Nanoscience Center, Aarhus University, Denmark

*Correspondence*
Mads Sloth Vinding, Center of Functionally Integrative Neuroscience, Aarhus University, Nørrebrogade 44, 10G-5-31, DK-8000 Aarhus C, Denmark

Main text word count: 2802
Abstract word count: 206
4 Figures, 1 Table

**ABSTRACT**

**Purpose:** Some advanced RF pulses, like multi-dimensional RF pulses, are often long and require substantial computation time due to a number of constraints and requirements, sometimes hampering clinical use. However, the pulses offer opportunities of reduced-FOV imaging, regional flip-angle homogenization, and localized spectroscopy, e.g., of hyperpolarized metabolites. We propose a novel deep learning approach to ultra-fast design multi-dimensional RF pulses with intention of real-time pulse updates.

**Methods:** The proposed neural network considers input maps of the desired excitation region of interest, and outputs a multi-dimensional RF pulse. The training library is retrieved from a large image database, and the target RF pulses trained upon are calculated with a method of choice.

**Results:** A relatively simple neural network is enough to produce reliable 2D spatial-selective RF pulses of comparable performance to the teaching method. For binary regions of interest, the training library does not need to be vast, hence, re-establishment of the training library is not necessarily cumbersome. The predicted, single-channel pulses were tested numerically and experimentally at 3 T.

**Conclusion:** We demonstrate a relatively effortless training of multi-dimensional RF pulses, based on non-MRI related inputs, but working in an MRI setting still. The prediction time of few milliseconds renders real-time updates of advanced RF pulses possible.

## INTRODUCTION

Multi-dimensional RF pulses have several advantages, but clinically viable implementations are distinctly lacking. This is especially true for parallel-transmit, where the number of controls increase, (local) specific-absorption-rate constraints apply, and $B_{1+}$ maps are essential (1). 2DRF pulses are commonly computed for small tip angles (STA) with (magnitude) least-squares optimizations (2,3), and scaling of the pulse amplitudes enable reasonably good large-tip-angle (LTA) pulses (4). Alternatively, optimal control (OC) theory is often used in genuine LTA pulse designs (1,5–7). These iterative procedures can handle many, constrained controls, i.e., the (long) train of pulse amplitudes and phases, and naturally impart a compromise between fidelity and computation time (1,6).

Deep learning (DL) and convolutional neural networks (NN) experience a major interest in general these years, due to significant prediction and equipment improvements. DL and NNs have also advanced into the field of MRI. Yet on RF pulse optimization, machine learning-type methods have to our knowledge only been exploited in a very limited number of cases, e.g., for RF shim weight prediction as proposed by Ianni et al. (8).

In this study, we propose DL as a novel procedure for generation of multi-dimensional RF pulses. We were inspired by Refs. (9,10), where an image databases, e.g., Imagenet (11), facilitate training of NNs for image reconstruction. We are interested in 2DRF pulses for reduced-FOV imaging (12), and related to dissolution-dynamic-nuclear-polarization applications (13); arterial spin labeling (14); and multiband EPI (15). The proposed method we present here, however, is general and not limited to those applications.

## METHODS

Here, a 2DRF pulse derived from a DL NN is denoted a DL-predicted pulse. For supervised DL, an appropriate *input dataset* (e.g., excitation profiles) and a corresponding *target dataset* (i.e. 2DRF pulses produced by a *target method* (TM)) are needed.

Traditional pulse designs, e.g., OC methods, are typically evaluated, during optimization, by how well the actual and desired magnetizations match and perhaps by how the pulses confine to limits. However, the NNs here are evaluated, during training, by how well the predicted and target pulses match. Hence, the magnetization response is information we use after DL as indirect, secondary evaluation.

The input and target datasets together constituting a *library* are commonly split into three sub libraries, used for updating NN fitting coefficients (*train*, 60%), tuning and monitoring for over-fitting (*validation*, 20%), and a final DL-prediction evaluation by an unseen library (*test*, 20%).

The proposed workflow is shown in Figure 1a.

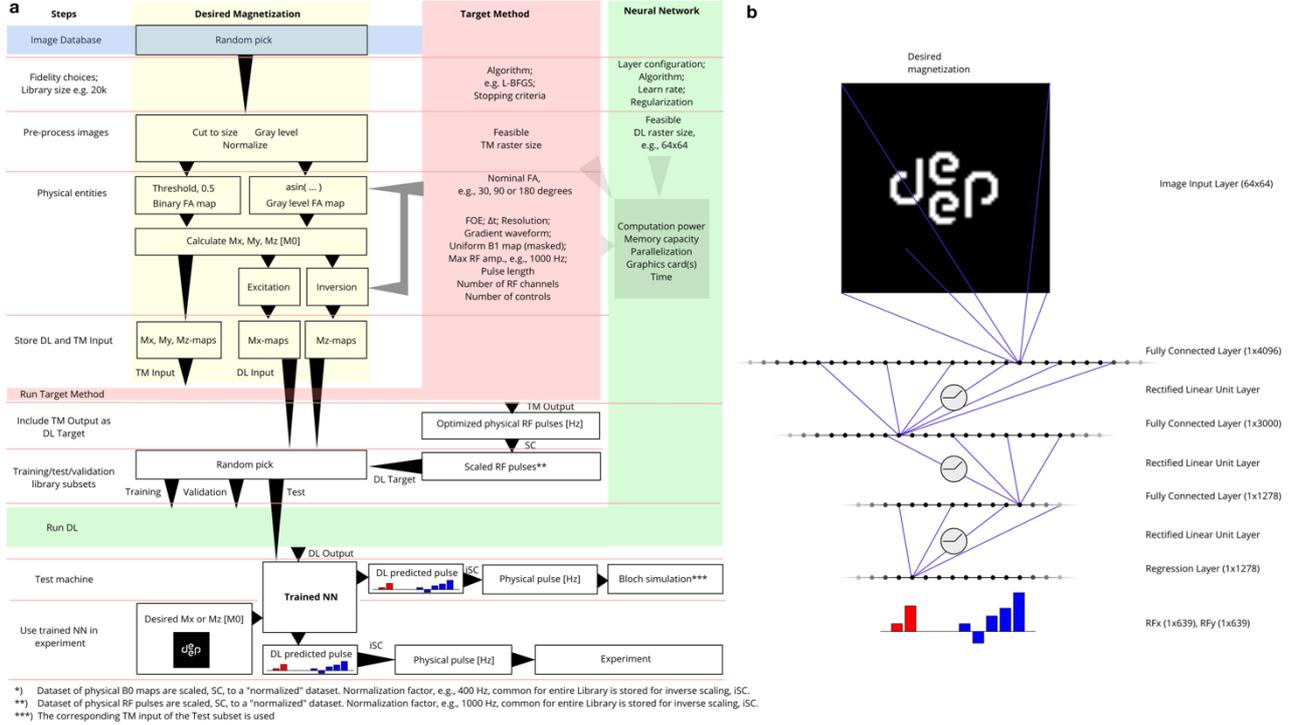

Figure 1. (a) Workflow diagram of proposed method. We have included scaling (SC) of physical entities to normalized variables, and the inverse scaling (iSC) as a mean to avoid very distinct numerical differences between input and target datasets. (b) NN construction diagram. Each dot in a fully connected layer is connected to all other dots in adjacent layers (not shown for clarity). Each connection between the fully connected layers are filtered by a rectified linear unit (not shown for clarity). The NNs work for one particular gradient waveform, and must be adjusted if another waveform is needed.

**Target Methods**

We investigated target datasets from two TMs implemented in blOCh (16): the OC LTA method of Ref. (1), denoted $TM_{LTA}$; and the regularization approach of Ref. (17) implemented as an STA method, denoted $TM_{STA}$. Specific optimization parameters are given in Table 1. The $TM_{LTA}$ enforced a hard constraint separately on *x*- and *y*-peak amplitudes of 1 kHz, while the $TM_{STA}$ applied Tikhonov regularization after finding the optimal regularization through an L-curve approach.

We used a spiral excitation k-space trajectory based on Ref. (18) obeying gradient amplitude and slew-rate constraints that were set to 40 mT/m and 180 T/m/s, respectively. The field of excitation was in all cases 25 cm, and the spatial grid was 64 by 64 in the axial plane. The resulting 16-turn, 6.39-ms long gradient waveform was rasterized into 10-µs steps.

**Training Libraries**

Input datasets were processed by randomly picking images from the Imagenet database (11). Images were cut to 64 by 64 pixels, gray scaled, and normalized. The input datasets were, depending on the library type, characterized by 1) pattern profiles and 2) nominal flip angles (FA), being 30°, 90°, or 180°. The pattern profiles could either be gray scale (Gr) or black/white (BW), with the latter formed by a 0.5-threshold. Black and white would correspond to 0° and nominal FA, respectively. Gray levels were converted by arcus sinus to designated FAs, adjusted such that the nominal FA was assigned to white pixels. For the 30° and 90° excitation cases, the input dataset would reflect the $M_x$ magnetization, i.e., with 0°, 30° and 90° corresponding to values 0, 0.5, and 1, respectively. For the inversion cases, the input dataset reflected the $M_z$ magnetization, i.e., with 0° and 180° corresponding to 1 and -1, respectively. All pattern profiles were masked with a super ellipsis shape.

We established eight libraries, see Table 1. Libraries 1, 2, 4, 5, 7, and 8 included 20,000 (20k) cases. Library 3 consisting of library 1 and 2 thus included 40k cases. Library 6 included 8k cases.

The target outputted by the TMs were arranged in arrays of length 1278, normalized to the largest value, with the normalization factors stored for back-conversion.

To assess the significance of the *library size*, we randomly split libraries 1 and 4 into smaller subsets of sizes 1k, 2k, 3k, 4k, 5k, 7.5k, 10k, 12.5k, 15k.

For libraries 5+7, and 8 the final FA was achieved by an optimized upscaling of libraries 4 and 1, respectively. The optimization involved stepping through a number of RF scaling factors and picking the one minimizing the normalized root mean square error (NRMSE) of the magnetization profile from a Bloch simulation against the desired scaled pattern.

Table 1: Libraries, desired magnetizations, target methods and deep learning parameters used in the workflow and neural network of Figure 1a and 1b, respectively.

| Library | | Desired Magnetization | | Target Method | | | | | | Deep Learning | | | |
|---|---|---|---|---|---|---|---|---|---|---|---|---|---|
| Reference* | Size | Pattern | FA [°] | TM | Scaling# | Max iterations@ | RF hard constraint | Tikhonov regularization& | Tolerances$ | Epochs± | Minibatch size | Learn rate | L2 regularization |
| 1 | 20k | BW | 90 | LTA | dna | 1000 | 1 kHz | dna | 5e-6/5e-6 | 1000 | 30 | 3e-3 | 1e-4 |
| 1b | 15k | - | - | - | - | - | - | - | -/- | - | - | - | - |
| 1c | 12.5k | - | - | - | - | - | - | - | -/- | - | - | - | - |
| 1d | 10k | - | - | - | - | - | - | - | -/- | - | - | - | - |
| 1e | 7.5k | - | - | - | - | - | - | - | -/- | - | - | - | - |
| 1f | 5k | - | - | - | - | - | - | - | -/- | - | - | - | - |
| 1g | 4k | - | - | - | - | - | - | - | -/- | - | - | - | - |
| 1h | 3k | - | - | - | - | - | - | - | -/- | - | - | - | - |
| 1i | 2k | - | - | - | - | - | - | - | -/- | - | - | - | - |
| 1j | 1k | - | - | - | - | - | - | - | -/- | - | - | - | - |
| 2 | 20k | Gr | - | - | - | - | - | - | -/- | - | - | - | - |
| 3 (1, 2) | 40k | BW+Gr | - | - | - | - | - | - | -/- | - | - | - | - |
| 4 | 20k | BW | 30 | STA | - | dna | dna | logspace(-6,10,25) | dna | - | - | - | - |
| 4b | 15k | - | - | - | - | - | - | - | - | - | - | - | - |
| 4c | 12.5k | - | - | - | - | - | - | - | - | - | - | - | - |
| 4d | 10k | - | - | - | - | - | - | - | - | - | - | - | - |
| 4e | 7.5k | - | - | - | - | - | - | - | - | - | - | - | - |
| 4f | 5k | - | - | - | - | - | - | - | - | - | - | - | - |
| 4g | 4k | - | - | - | - | - | - | - | - | - | - | - | - |
| 4h | 3k | - | - | - | - | - | - | - | - | - | - | - | - |
| 4i | 2k | - | - | - | - | - | - | - | - | - | - | - | - |
| 4j | 1k | - | - | - | - | - | - | - | - | - | - | - | - |
| 5 (4) | 20k | - | 30→90 | - | [1.5:0.05:4] | - | - | logspace(-6,10,25) | - | - | - | - | - |
| 6 | 8k | - | 180 | LTA | dna | 1000 | 1 kHz | dna | 5e-6/5e-6 | 3000 | 100 | - | - |
| 7 (4) | 20k | - | 30→180 | STA | [4:0.05:10] | dna | dna | logspace(-6,10,25) | dna | - | - | - | - |
| 8 (1) | 20k | - | 90→180 | LTA | [1:0.05:4] | 1000 | 1 kHz | dna | 5e-6/5e-6 | - | - | - | - |

*: B (A) means library B is developed from library A through scaling #. Cd means library Cd is a subset of library C.
-: same as above;
dna: does not apply;
#: The RF scaling factor producing the lowest NRMSE was found in this array specified as [C:D:E] i.e. of linearly spaced points between C and E in steps of D.
@: The TM$_{LTA}$ (L-BFGS) will stop after this number of iterations even if tolerances, $, are not met.
$: X/Y are stopping thresholds of X change of controls and Y change of cost function, i.e., changes of either parameter below their threshold would characterize the optimization as converged.
&: Optimal Tikhonov regularization was found through the L-curve method by sweeping through logspace(A,B,N), i.e. an array of N logarithmically spaced points between decades $10^A$ and $10^B$, and automatically picking the corner value.
±: number of epochs was assured to be high such that convergence was qualitatively observed.

## Neural Networks and Deep Learning

The NN construction, see Figure 1b, consisted only of an image input layer (size 64x64); three fully connected layers of sizes 4096, 3000, and 1278; and rectifier linear unit layers in between; and lastly a regression layer of size 1278, which constitutes the output and matches the size of the target.

The DL was done with the stochastic-gradient-descent-with-momentum algorithm in MATLAB 2018a (Mathworks, Natick, MA). Parameters like number of epochs, $L_2$ regularization, minibatch sizes, learn rate etc. are tabularized in Table 1. The parameters were investigated by starting from the MATLAB default value and if need be adjusted until a reasonable convergence was observed. Hence, our DL success criteria were elimination of overfitting and establishing convergence, and otherwise to use equal parameters for the NNs we compared directly.

The DL was run on a workstation with an NVIDIA Tesla P100 16GB GPU.

We generally use peak amplitudes and the NRMSE to evaluate performance of each trained NN by comparing actual and desired magnetizations derived from DL-predicted and TM-calculated pulses belonging to the test subset and with exemplar demonstrations. With the library size assessment, we also compare the NRMSE of DL-predicted pulses against TM-calculated pulses. For statistical assessment, we employed the Wilcoxon rank-sum test.

**Phantom and In vivo Experiments**

A 10k BW-type NN, i.e. reference 1d in Table 1 was used in experiments.

We implemented the 2DRF pulses into a spin-echo sequence on our Siemens PrismaFit 3T system, using Pulseq (19). Acquisition parameters: 256 lines, FOV 25 cm, 5 mm axial slice (isocenter), TE 20 ms, TR 1s, 2 acquisitions, and duration 8:32. We did phantom measurements with an oil-containing ball of approximately 25 cm diameter. We did two scans, where the BW-type, desired magnetization was the word "deep" with 1) the TM-calculated pulse, and 2) the DL-predicted pulse. We then did a scan with a Gr-type, desired magnetization. In vivo experiments were conducted with a healthy male volunteer, who gave informed consent prior to participation. We did one scan with a DL-predicted pulse of a square excitation.

# RESULTS

## Training libraries

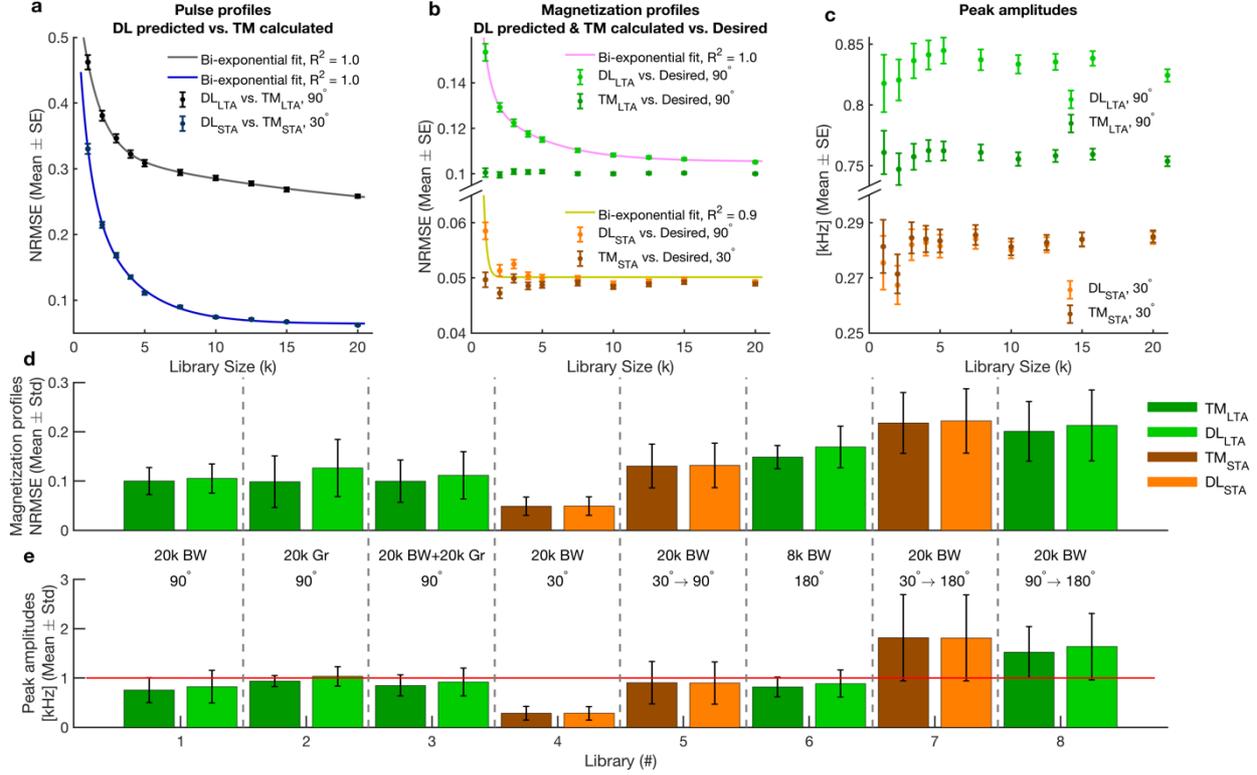

Figure 2. Performance and pulse metrics deduced from the test subsets of libraries 1 to 8. (a) NRMSE of DL-predicted pulses against TM-calculated pulses of libraries 1-1j and 4-4j. (b) NRMSE of magnetization profiles from DL-predicted and TM-calculated pulses (with respect to their desired magnetization patterns) of libraries 1-1j and 4-4j. (c) Peak amplitudes of DL-predicted and TM-calculated pulses of libraries 1-1j and 4-4j. (d) NRMSE of magnetization profiles of libraries 1 to 8. (e) Peak amplitudes of libraries 1 to 8. (a-c) Show mean values with error bars being standard error (SE). (d, e) Show mean values with error bars being standard deviation (Std).

The curve-fit parameters of (a-c) are presented in the Supporting Information. The red line in (e) signify the hard limit applied in $TM_{LTA}$. In (d), DL NRMSE values were significantly higher than the corresponding TM values, i.e. pairwise, for all libraries except with libraries 4 and 5, where there was no significant difference. In (e), DL peak amplitudes were significantly higher than the corresponding TM values, i.e. pairwise, for all libraries except with libraries 4, 5, and 7, where there was no significant difference. TM and DL peak amplitudes were significantly below 1 kHz in libraries 1, 3 to 6; and significantly above in libraries 7 and 8. For library 2, the peak amplitudes of the $TM_{LTA}$ and $DL_{LTA}$ were significantly below and above 1 kHz, respectively, but the $DL_{LTA}$ was not significantly different from 1.03 kHz. NRMSE and peak amplitudes of $DL_{STA}$ of libraries 5 and 7 are significantly larger than those of $DL_{LTA}$ of libraries 1 and 6, respectively.

Figure 2 shows the performances of each library we tested. On the library size assessment, pulse and magnetization data, with mean NRMSE values with standard errors (SE) are shown in Figure 2a and 2b, respectively. The bi-exponential curve-fit data are presented in the Supporting Information. While the $DL_{STA}$ appears converged (Figure 2b) the $DL_{LTA}$ can reach within 0.5% NRMSE off the 20k $TM_{LTA}$ average value (of 0.1 NRMSE) with a 34k-sized library. From this point on, in this setup, increasing the library size has no significant influence.

Figure 2c shows the peak amplitudes (mean and SE). The TM-calculated peak amplitudes depend on the desired magnetization, where larger areas or more difficult shapes presumably require higher amplitudes - for $TM_{LTA}$ often at the limit. The $DL_{LTA}$-predicted amplitudes are on average higher, but well within the $TM_{LTA}$ limit of 1 kHz. The $DL_{STA}$ and $TM_{STA}$ peak amplitudes are for the larger libraries essentially the same.

Figure 2d and 2e show the NRMSE and peak amplitudes of libraries 1 to 8, respectively. Comparing library 1 and 5, 90° BW-types, we see significantly lower NRMSEs and peak amplitudes for the LTA method than the STA method of 30° STA pulses scaled to 90° before training.

On inversion pulses, libraries 6 to 8, we see the significantly best performance of the LTA, for designated 180° pulses (even with a relatively smaller library), when compared to 90° pulses scaled to 180° and 30° STA pulses scaled to 180°. Scaling results in larger variations too.

The significance of libraries 1 to 4 on the desired magnetization pattern type (BW and Gr) are considered with the examples shown in Figure 3 and Supporting Figure S1. Expanding the BW-type library improves both BW- and Gr-type patterns (Figure 3, columns 3 to 5, rows 1 to 4), but for 90° pulses especially, the latter type is best achieved with, at least, a Gr-type library (Figure 3m), but overall preferentially with both BW- and Gr-type examples (Figure 3n).

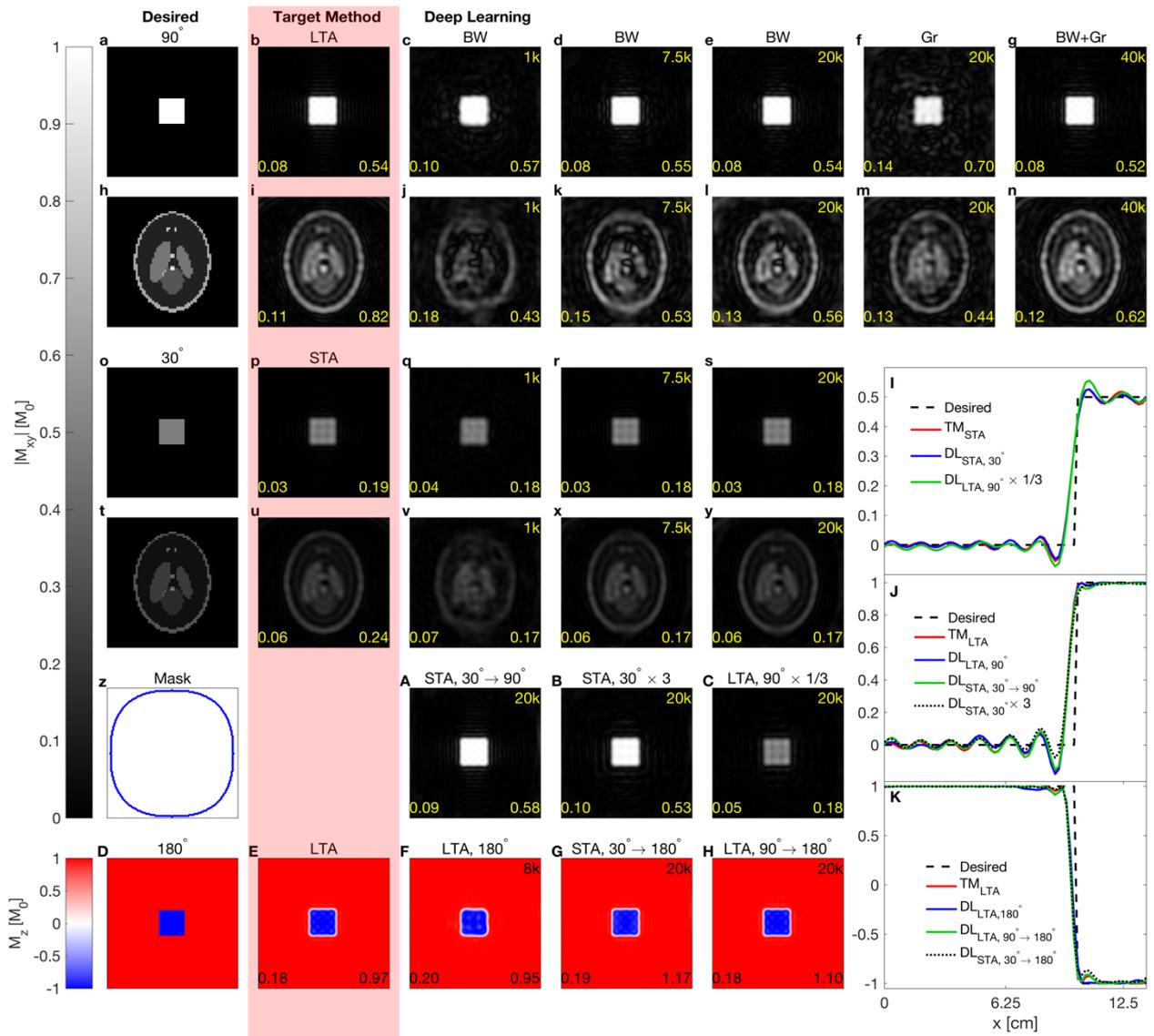

Figure 3. Bloch simulation example demonstrations of the different libraries. Column one shows the desired magnetization patterns (a, h, o, t, D) and the mask boundary distinguishing a "don't care region" (outer section) from the "region of interest" (inner section). Column two shows TM performance (b, i, p, u, E) given the desired patterns of column one. Printed numbers in the lower left and right corners are NRMSE and peak amplitude [kHz], respectively. Column three and on, show NN performance of different library types (specified above each column). Printed numbers in the upper right corners are the library sizes. For example, (j-l) increasing performance with the Gr-type pattern going from 1k to 7.5k to 20k using a BW-type NN. (f, m) a Gr-type NN; (g, n) a NN trained with both Gr- and BW-type patterns. (A) library 5, 30° $TM_{STA}$ pulses scaled to 90° before DL. (B) library 4, 30° $TM_{STA}$, where the image input corresponding to (o) is multiplied by 3 to produce a 90° excitation, hence, targeting (a). (C) library 1, 90° $TM_{LTA}$, where the image input corresponding to (a) is divided by 3 to produce a 30° excitation, hence, targeting (o). (I-K) Profiles from half of the central horizontal traces of specified examples.

Figure 3A (library 5) compares to Figure 3e (library 1). Hence, it is indeed possible to train 90° pulses with an STA method. Figure 3B and 3C demonstrate up and down scalability of the NNs, respectively. For example, the $DL_{STA}$ NN associated with Figure 3B expects a BW image input of 0 and 0.5, where the FA is supposed to be 0° and 30°, respectively. If the pixels of value 0.5 are replaced with value 1.5, the NN outputs 90°, and so forth. This is, as Figure 3B and 3C show, to similar performances of the corresponding Figure 3e and 3s, respectively.

The inversion examples, Figure 3, row 6, show quite good and comparable performance between the different NNs, having in mind the smaller size of the designated 180° library actually constrained in the $TM_{LTA}$ method to the same level, 1 kHz, as the 90° pulses.

Specified trace profiles of Figure 3I to 3K highlight comparable ripple-tendencies of the presented cases.

**Phantom and In vivo Experiments**

Figure 4, shows the numerical and experimental results of the $DL_{LTA}$-predicted and the $TM_{LTA}$-calculated pulses. The first two rows are the phantom results. In Figure 4d we show horizontal image traces. We note the signal suppression in this example is better for the $DL_{LTA}$-predicted than for the $TM_{LTA}$-calculated pulse.

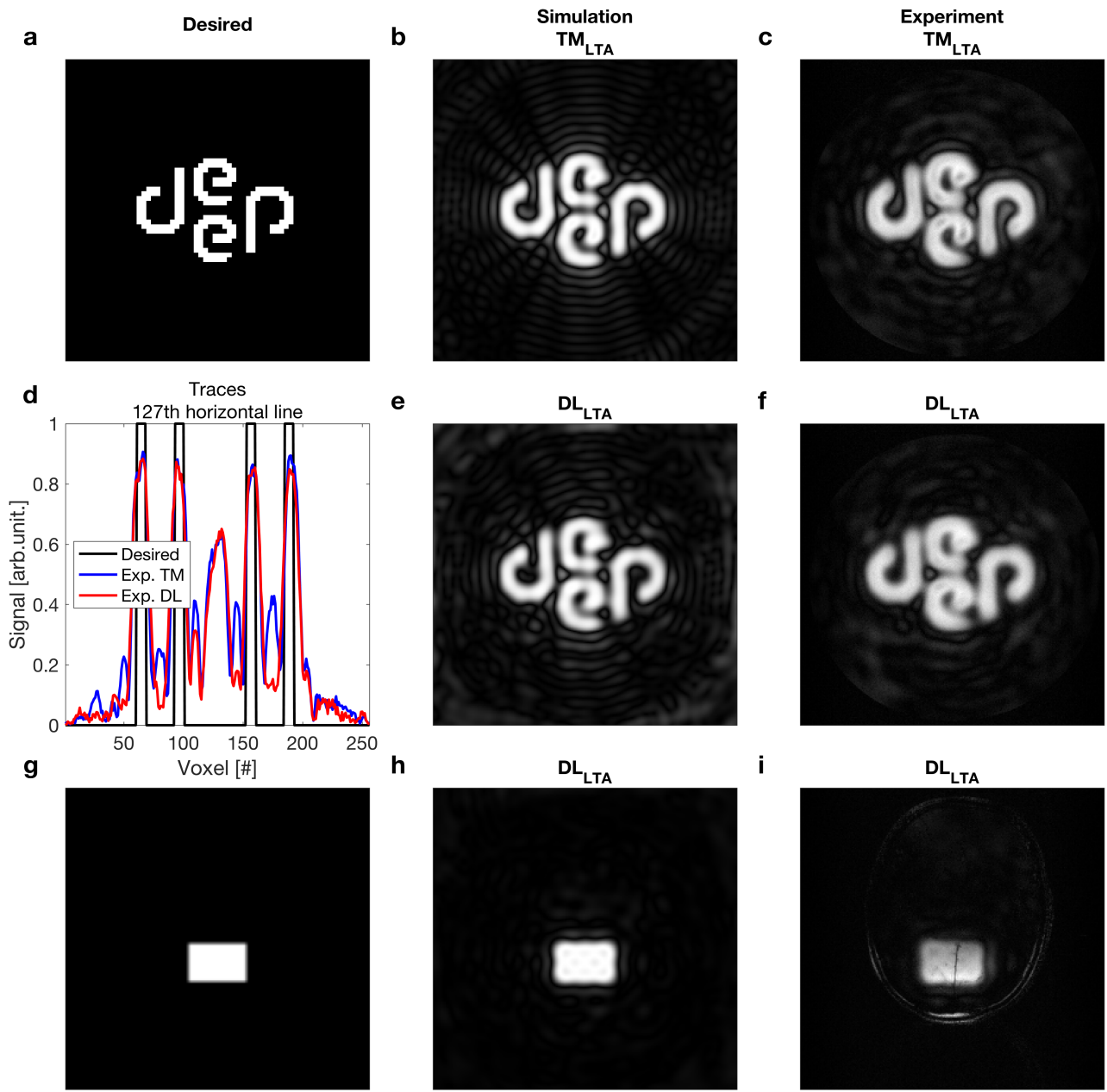

Figure 4. Experimental phantom and in vivo results. (a, g) Desired magnetization patterns, 90°. (b, e, h) Simulations of transverse magnetization magnitudes. (c, f, i) Experimental results. (d) 127th horizontal trace lines through (a), (c), and (f).

The third row of Figure 4 shows our in vivo results. As evident, there is a suboptimal suppression of lipid signal near the skull, which is typically seen in reduced-FOV sequences not incorporating fat-suppression as the case here (20,21). We have shown the performance of a Gr-type, desired magnetization with the BW-type NN in Supporting Figure S2.

Prediction time was for the test subset 7.2 ms on average.

**DISCUSSION**

We have proposed an on-the-fly prediction method for designing multi-dimensional RF pulses based on a desired excitation pattern by DL NNs. As few as one to two thousand training cases, perhaps fewer, is especially for the $TM_{STA}$, enough to see, feasible learning results and pulses that actually perform multi-dimensional excitation similar to what we usually see with calculated pulses bases on models of magnetization dynamics etc. We have demonstrated that spatially varying excitation patterns, which we referred to as Gr, can be achieved to some extend with NNs trained with only BW patterns. We could improve spatial variation, by tailoring Gr-library number 2, and further extend this to library 3. The Gr excitation patterns, are for DL and TM more demanding in terms of peak amplitudes, than the BW excitation patterns, see Figure 2. Given the lack of RF control by our DL predictions, we will in future projects investigate solutions for this.

Up to the point up-scaling did not take place, e.g., from 30° or 90° to 180°, and except for the Gr $DL_{LTA}$ case, the peak amplitudes were of the test subsets significantly below 1 kHz that was enforced as a limit in the $TM_{LTA}$ cases.

Means to further improve control of RF power could be by handles in the DL, perhaps the $L_2$-regularization, or by further limiting the TM pulses. That is, establishing a relation between TM hard limits and prediction tendencies, i.e., a safety overhead regularization. Finally, one can always check the DL-predicted pulses, and if necessary adjust either the desired magnetization, and predict a new one; downscale or clip the DL-predicted pulse at the expense of flip-angle reduction and/or minor excitation distortions; or, if the pulse sequence and application allows, apply the VERSE principle and decrease RF amplitudes for a prolonged gradient waveform (22).

Compared to $L_2$-regularization, learning step sizes, number of training epochs, and how to actually construct the NN skeleton, we were mostly concerned with the library size. As building a library can be cumbersome, but necessary every time, e.g., a new gradient waveform is needed, we wanted to get a feeling of how many training examples, we should expect to produce with the TM. Our library size comparison shows that it may depend on the desired magnetization. Simple shapes like a square may require just few thousand training examples, and as mentioned above difficult gray-scale patterns require more. We have shown that $DL_{LTA}$-predicted pulses require vast libraries, when comparing to the $TM_{LTA}$ calculated pulses. We have indicated convergence trends, when comparing to the $TM_{LTA}$, but these depend on many parameters. There is likely room for improvement and earlier convergence to the $TM_{LTA}$ than these trends forecast now. For instance, there may be a compromise between even better $TM_{LTA}$ pulses and the size of the library.

At times, when investigating the $L_2$-regularization, learning step sizes, and the NN skeleton, we saw tendencies of overfitting, which can be overcome by several means (23–25), e.g., harder $L_2$-regularization, drop-out layers, and larger samples. We chose in this study to present the most basic NN that we could make work for us. Our future studies will concern improving the NN. For example, by DL of desired and/ or simulated magnetizations (see Supporting Information), and deducing the pulse from the hidden layers, bearing resemblance to OC (1,26,27).

In the Supporting Information, we have included preliminary data of $B_0$-inhomogeneity compensation achieved by, e.g., extending the third dimension of the Input dataset from 64x64x1 to 64x64x2, incorporating a $B_0$ map in the second layer. By the same approach, we plan to include $B_{1+}$ maps in future studies. 3DRF will likely demand 4D inputs or mechanisms that link multiple parameters belonging to the same voxel together.

We propose to use DL-predicted pulses as initial guesses for subsequent optimizations. A good initial guess can be essential for some applications (1). We also propose to use DL for pulses designed with time-demanding or advanced techniques. This could be quantum mechanical approaches on multi-spin systems (7,28,29), applications with very precise control updates (30), or certain characteristics like pulse envelope smoothness (31).

With the presented NNs, predicting a pulse takes on average about 7 ms on a laptop computer. For reference, a Bloch simulation of the pulse magnetization response in our simulation environment takes about 300 ms. The $TM_{STA}$ and $TM_{LTA}$ spend approximately 1.5 s (incl. regularization assessment (17)) and several minutes (depending on parallelization, and several factors (1)) per pulse, respectively. The short DL-prediction time and even with a slower Bloch simulation for visualization opens up for actual real-time ROI drawing and pulse design, a goal we have pursued before (26,27). In the Supporting Information Video S1, we have included a movie demonstrating how such an event could play out. Alternatively, to manually drawing a ROI, one could use a segmentation strategy (32,33).

Our proposed solution to pulse design bears resemblance to MR Fingerprinting (34), where a significant scan time burden associated with quantitative MRI has been replaced with fast spatially and temporally incoherent scans, followed by multi-parametric mappings through a look-up in an offline-generated table. It is possible to train NNs to a variation of setups and purposes offline with practically unlimited time, enabling fast workflows without much sacrifice in scan performance or duration.

The proposed method will be accessible at github.com/madssakre/DeepControl (35).

**CONCLUSION**

A, to the best of our knowledge, novel, and ultra-fast approach for multi-dimensional RF pulse design has been presented, and demonstrated with phantom and in vivo experiments. While clinical applications cannot wait too long for results, this real-time pulse design tool, is prepared and enriched offline from the scanner with a method of choice and all the complexity that one wishes to include. The deep learning framework is flexible and we have shown an easy starting point for developing more sophisticated pulse designing neural networks, and we will pursue new opportunities for neuroscientific applications and with clinical applicability in mind.

**ACKNOWLEDGEMENT**

We acknowledge support from VILLUM FONDEN, Harboefonden, Kong Christian den Tiendes Fond, and Centre for Scientific Computing, Aarhus.

**FIGURE CAPTIONS**

**Supporting Information**

Supporting Information legend: Supporting data for training libraries, experimental results, and $B_0$ inhomogeneity compensation.

Supporting Figure S1. All library sizes, transverse magnetization profile, examples for corresponding to Fig 3 in the main text. Columns from left to right correspond to the library sizes of 1k, 2k, 3k, 4k, 5k, 7.5k, 10k, 12.5k, 15k and 20k. Rows 1 and 2 are made with $TM_{LTA}$ method with FA = 90º, and rows 3 and 4 are with the $TM_{STA}$, FA = 30º. The training libraries were of the BW-type.

Supporting Figure S2. Phantom results. a) Desired Gr-type, desired magnetization pattern; b) $|M_{xy}|$ simulation of TM-calculated pulse; c) $|M_{xy}|$ simulation of the $DL_{LTA}$-predicted pulse; d) experimental result of the $DL_{LTA}$-predicted pulse; e) the horizontal center-traces through images in a,b,c,d).

Supporting Figure S3. (a) Workflow diagram of the proposed method with $B_0$-inhomogeneity compensation. We have included scaling (SC) of physical entities to normalized variables, and the inverse scaling (iSC) as a mean to avoid very distinct numerical differences within and between input and target datasets. (b) NN construction diagram. Each dot in a fully connected layer is connected to all other dots in adjacent layers (not shown for clarity). Each connection between the fully connected layers are filtered by a rectified linear unit (not shown for clarity).

Supporting Figure S4. (a) Simulations and field maps. Upper left (Desired magnetization pattern, 30º). Row 1 from column 2 to 12: nominal $B_0$ map followed by scaled ones. Column 1 simulations are with an ideal $B_0$ map of zero off resonance. Row 2: non-compensated $TM_{STA}$ pulses. Row 3: non-compensated $DL_{STA}$ pulses. Row 4 to 8: compensated $DL_{STA}$ pulses with increasing library sizes. Row 9: compensated $TM_{STA}$ pulses. (b) NRMSE values for magnetizations against the desired magnetization (upper left in (a)).

Video S1. Movie showing a real-time pulse design event. A mouse cursor is used to draw squares on top of an anatomical map and each mouse click triggers a 2DRF pulse prediction. The desired pattern, the predicted pulses and the Bloch simulation are shown updated in real time.